\documentclass[%
 reprint,
superscriptaddress,
nofootinbib,
 amsmath,amssymb,
 aps,
]{revtex4-1}

\usepackage[colorlinks]{hyperref}
\usepackage{graphicx}
\usepackage{dcolumn}
\usepackage{bm}
\usepackage{color}
\usepackage{placeins}

\definecolor{red}{rgb}{0.9, 0,0}
\definecolor{cerulean}{rgb}{0., 0.62,0.9}
\definecolor{navy}{rgb}{0.05, 0.05,0.8}

\begin{document}

\title{A Trigger for Displaced Muon Pairs Following the CMS Phase II Upgrades}

\author{Yuri  Gershtein}
\affiliation{Department of Physics and Astronomy, Rutgers University, Piscataway, NJ 08854, U.S.A.}

\author{Simon Knapen}
\affiliation{School of Natural Sciences, Institute for Advanced Study, Princeton, NJ 08540, U.S.A.}

\date{\today}

\begin{abstract}
We show that the phase II upgrade of the CMS tracking detector could enable the experiment to trigger on very low mass $\mathcal{O}(1\,\text{GeV})$ displaced muon pairs with minimal $p_T$ cuts. As a result, CMS can be competitive with LHCb when searching for low mass displaced exotics originating from heavy flavor decays. The method can also be applied to signatures without muons but with a moderate amount of MET, $H_T$ or multiple displaced vertices in the event. 
\end{abstract}

\maketitle

\section{\label{sec:intro}Introduction}
With the LHC now gearing up for its high luminosity phase, all detectors will undergo substantial upgrades to cope with the expected high levels of pile-up. The tracking detectors especially are an important component of this effort. ATLAS \cite{Collaboration:2285585}, CMS \cite{CMSCollaboration:2015zni} 
and LHCb \cite{Bediaga:1443882} are all aiming to incorporate tracking information as early on in the trigger decisions as possible, for the phase II upgrade or sooner.  For CMS in particular, as of run 4, the modules in the outer tracker will consist of two closely spaced sensors, such that a rudimentary momentum measurement will be possible on a single module using the pairs of hits (stubs) \cite{Collaboration:2272264}. A momentum preselection of $p_T>$ 2 GeV on these track stubs then greatly reduces the occupancy, rendering preliminary track finding feasible under the stringent time constraints of the L1 trigger.

It has been shown recently that this approach can in fact reconstruct tracks with a transverse impact parameter up to roughly 10 cm, opening up the exciting possibility of a dedicated L1 trigger for long-lived particles (LLPs) \cite{Gershtein:2017tsv}. Concretely, \cite{CMS-PAS-FTR-18-018} showed that demanding a number of displaced tracks combined with a moderate $H_T$ requirement can greatly increase the trigger efficiency for Higgs bosons decaying to a pair of displaced dijets. In this work we build on \cite{Gershtein:2017tsv,CMS-PAS-FTR-18-018} by exploring the feasibility of a Level-1 trigger for a low mass displaced dimuon \mbox{\emph{vertex}} with minimum muon $p_T$ thresholds. This would give CMS greatly improved access to e.g.~exotic B-meson decays to light hidden sector particles, which can arise for instance from a light scalar mixing with the Higgs boson. In particular, we will show in Fig.~\ref{fig:yield} that this capability could make CMS competitive with LHCb \cite{Aaij:2016qsm,Aaij:2015tna} for this class of signatures. Dark photon models are another application, and have been studied extensively already in the context of the (upgraded) LHCb detector \cite{Aaij:2017rft,Ilten:2016tkc,Ilten:2015hya}.

Our strategy can be summarized as follows: We use the simulation framework described in \cite{Gershtein:2017tsv} to reconstruct tracks from the stubs retained by the future CMS outer tracker at the L1 trigger, and hereby include resolution smearing and multiple scattering. We subsequently find the most compatible vertex and require that the reconstructed mother particle trajectory points back to the interaction point. The latter is needed to suppress the large rate of vertices formed by random crossings of fake tracks. Once a suitable vertex is identified, its tracks can be matched to activity in the muon detectors, which should suffice to suppress the rate to $\mathcal{O}(\text{kHz})$.

This paper is organized as follows: We describe simulation framework in Sec.~\ref{sec:mc} and our analysis and results in Sec.~\ref{sec:analysis}. We conclude and present an outlook in Sec.~\ref{sec:summary}. In Appendix~\ref{sec:turnon} we provide some additional figures which we hope could be useful to recast our results for different models. 

\section{\label{sec:mc}Toy detector simulation}
\subsection{Signal and background simulation}
The example signal we consider is that of a light particle ($\varphi$) produced through the $B\to X_s \varphi$ decay with $\varphi$ decaying to muons ($\varphi \to \mu^+\mu^-$) through a displaced vertex. Such a signal can arise for instance from a light scalar mixing with the Higgs boson, which itself is a feature in many UV complete models of beyond the Standard Model physics. A model of this kind typically predicts correlations between the lifetime, production rate and branching ratios of the new particle, however in this work we simply parametrize the yield in terms of the lifetime ($c\tau$) of $\varphi$ and its production rate, expressed in terms of $\text{Br}[B\to X_s \varphi]\times \text{Br}[\varphi \to \mu^+\mu^-]$, where $\text{Br}[B\to X_s \varphi]$ is the inclusive $B$-meson branching ratio. The signal was generated with \verb+pythia 8+~\cite{Sjostrand:2006za,Sjostrand:2014zea} and we assumed an inclusive $b$-$\bar b$ cross section of $500\,\mu$b \cite{Aaij:2016avz}. 

Two sources of background are considered. First, and largest, is the random crossings of fake (i.e.~comprised of the unrelated stubs) tracks. We assume the number of fake tracks averages 30 per event, with all five track parameters (see e.g.~\cite{trackparam}) ($q/p_T$, $\phi$, $d_0$, $\tan\lambda$, and $z_0$) uniformly distributed. This estimate is based on estimates of the expected occupancy of the trigger system \cite{CMSCollaboration:2015zni}, and was reproduced by CMS with a full simulation \cite{CMS-PAS-FTR-18-018}. Not all combinations of parameters produce tracks with observable stubs, mostly due to the anti-correlation between the $q/p_T$ and $d_0$  (See Fig.~\ref{fig:fake_pars} in Appendix \ref{sec:turnon}). We throw uniformly distributed track parameters (see Table \ref{tab:fpars}) until we get a track that is expected to have at least four observable stubs.  

\begin{table}[h]
    \centering
    \begin{tabular}{c|c|c}
\hline
track parameter & low range & high range \\
\hline
$q/p_T$ & -1/(2 GeV) & 1/(2 GeV)\\
$\phi$ & 0 & 2$\pi$\\
$d_0$ & -14 cm & 14 cm\\
$\tan \lambda$ & -6 & 6\\
$z_0$ & -15 cm & 15 cm\\
\hline
\end{tabular}
    \caption{Ranges used for simulation of fake tracks. Not all combination of parameters can produce four or more valid stubs. (See Fig.~\ref{fig:fake_pars} in Appendix \ref{sec:turnon}.) }
    \label{tab:fpars}
\end{table}

The second source of background is the the production of known long-lived particles, dominated by the $K_S$. We used
inclusive QCD events generated with \verb+pythia 8+~\cite{Sjostrand:2006za,Sjostrand:2014zea} to estimate it.

We neglect interactions of prompt particles with the 
material of the detector, as they are relatively rare and are not expected to produce candidates that point back to the primary vertex (PV). High energy photon conversions point back to the PV, but are trivial to remove based on the smallness of the opening angle. In fact, it is likely that the vertex selection cuts in Sec.~\ref{sec:analysis}
remove most of them, though this is difficult to estimate with our toy simulation alone. It is therefore possible that a requirement on the opening angle will be needed, which could degrade the signal efficiency for very low masses ($m_\varphi\ll 1$ GeV).

\subsection{Track reconstruction}
The toy Monte Carlo to estimate track reconstruction and resolution is similar to the one in \cite{Gershtein:2017tsv}, with several important additions.

The toy tracker has six perfectly cylindrical double layers covering $|\eta|<2.4$. Hits in the double layers produce a stub if the azimuthal offset between them is consistent with one from a prompt track with $p_T > 2$ GeV. The hits are smeared according to the expected resolutions \cite{Collaboration:2272264}. We also make a simple estimate of multiple scattering in the detector, assuming tracks change direction of travel when crossing a tracking layer by a random angle with mean of zero and Gaussian sigma of $4\cdot 10^{-4}/p_T$ \cite{Collaboration:2272264, Tanabashi:2018oca}. We do not consider the effects of small, non-gaussian tails to the scattering distribution, since they are unimportant for the signal, and for background they can be viewed as a small contribution to the fake displaced tracks. To account for scattering in the pixel detector, we add scattering layers corresponding to approximate positions of the pixel layers.

\begin{figure}[t]
\includegraphics[width=0.4\textwidth]{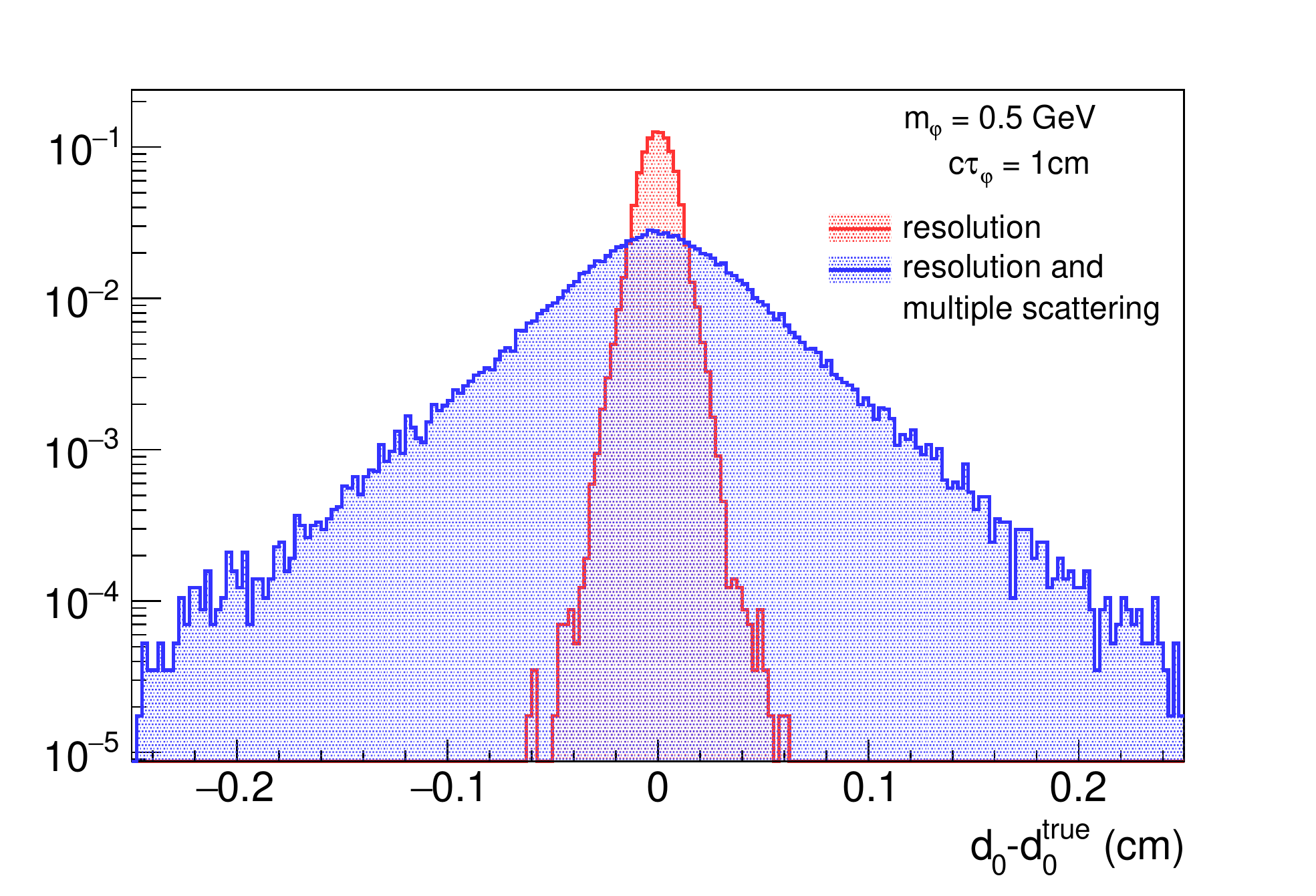}%
\caption{\label{fig:mcres_d0} Effect of multiple scattering on impact parameter resolution,  for a signal sample with $m_\varphi=0.5$ GeV and \mbox{$c\tau_\varphi=1$ cm.} For this signal sample, the median $p_T$ of $\phi$ is 5.8 GeV.}
\end{figure}

The resulting hits are then fit to a 5-parameter helical trajectory. Fig.~\ref{fig:mcres_d0} shows the impact of the multiple scattering on the track impact parameter resolution for a signal sample with $m_\varphi = 0.5$ GeV. The reconstruction efficiency depends on the required number of stubs, the $p_T$ and the transverse impact parameter. For example, for tracks with 4 stubs, the efficiency is roughly 80\%(50\%) for transverse impact parameter of 5cm (10cm). For plots of the efficiency in various cases we refer the reader to Ref.~\cite{Gershtein:2017tsv}.

\subsection{Vertex reconstruction}
The fitted tracks are the inputs to the vertex finding algorithm, which is deliberately kept as simple as possible, as proper vertexing is computationally prohibitive at the L1 trigger level. Our simplified vertex finder proceeds as follows: First we calculate the intersections between the two  helices in the transverse plane. If no intersection exist, we identify the transverse location of the candidate vertex with the point on the line defined by the centers of the circles for which the distance to the circle boundaries is equal. The distance between both circle boundaries measured along this line is a measure of the quality of the vertex in the transverse plane, and is labeled by $\Delta_T$. We subsequently compute the distance between the tracks in the $z$-direction for this candidate vertex, which we denote by $\Delta_z$. If two intersections are found, we calculate the distance in the $z$-direction between both tracks ($\Delta_z$) for each of the two solutions, and select the solution for which $|\Delta_z|$ is smallest. The $\Delta_T$ variable is set to zero in this case. 

\begin{figure}[t]\centering
\includegraphics[width=0.4\textwidth]{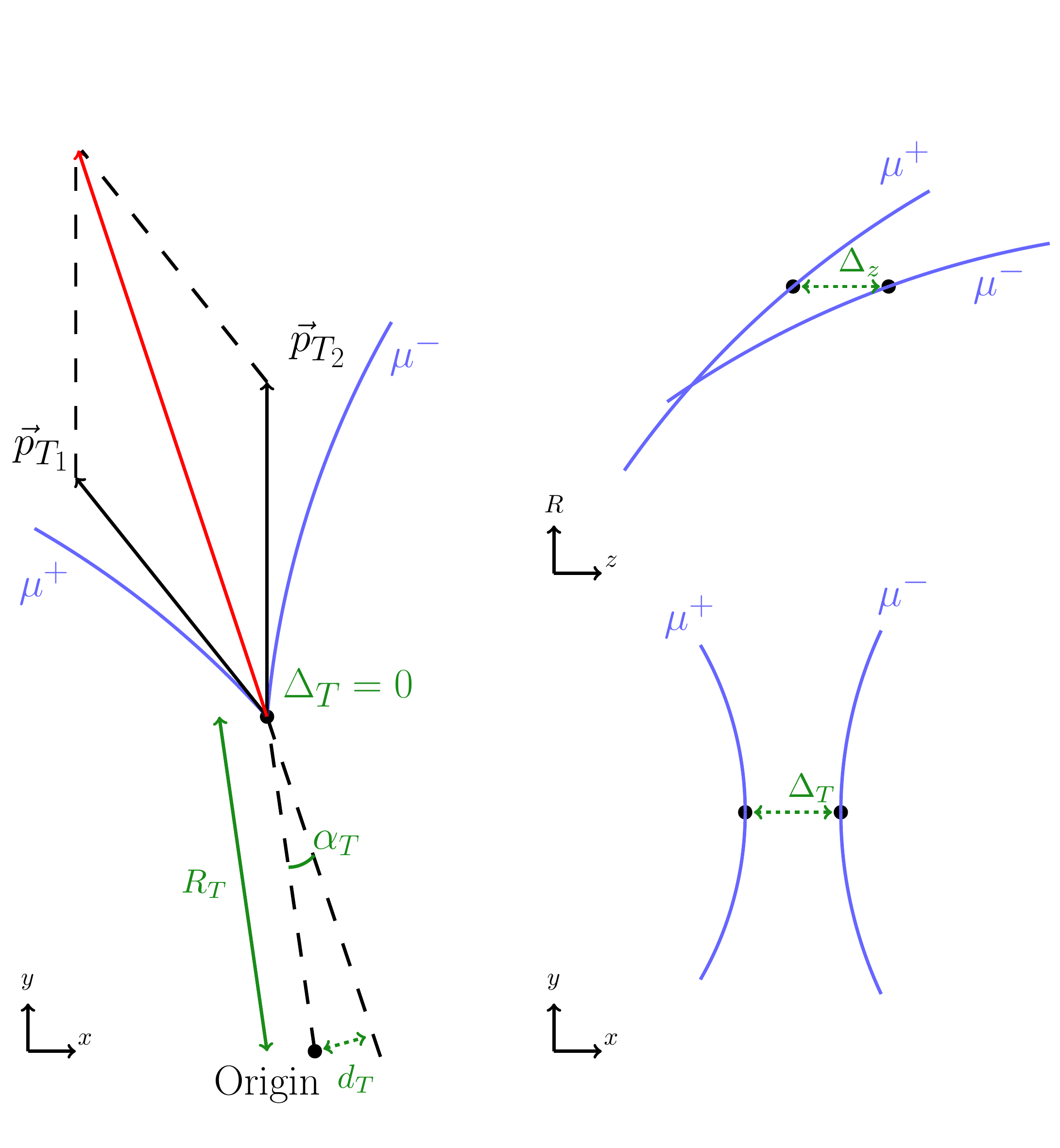}%
\caption{\label{fig:vertex_params} Most important vertex parameters. The left-hand diagram shows an example with $\Delta_T=0$ cm; the $\Delta_T>0$ cm case is shown in the bottom right diagram.}
\end{figure}

\begin{figure}[t]
\includegraphics[width=0.4\textwidth]{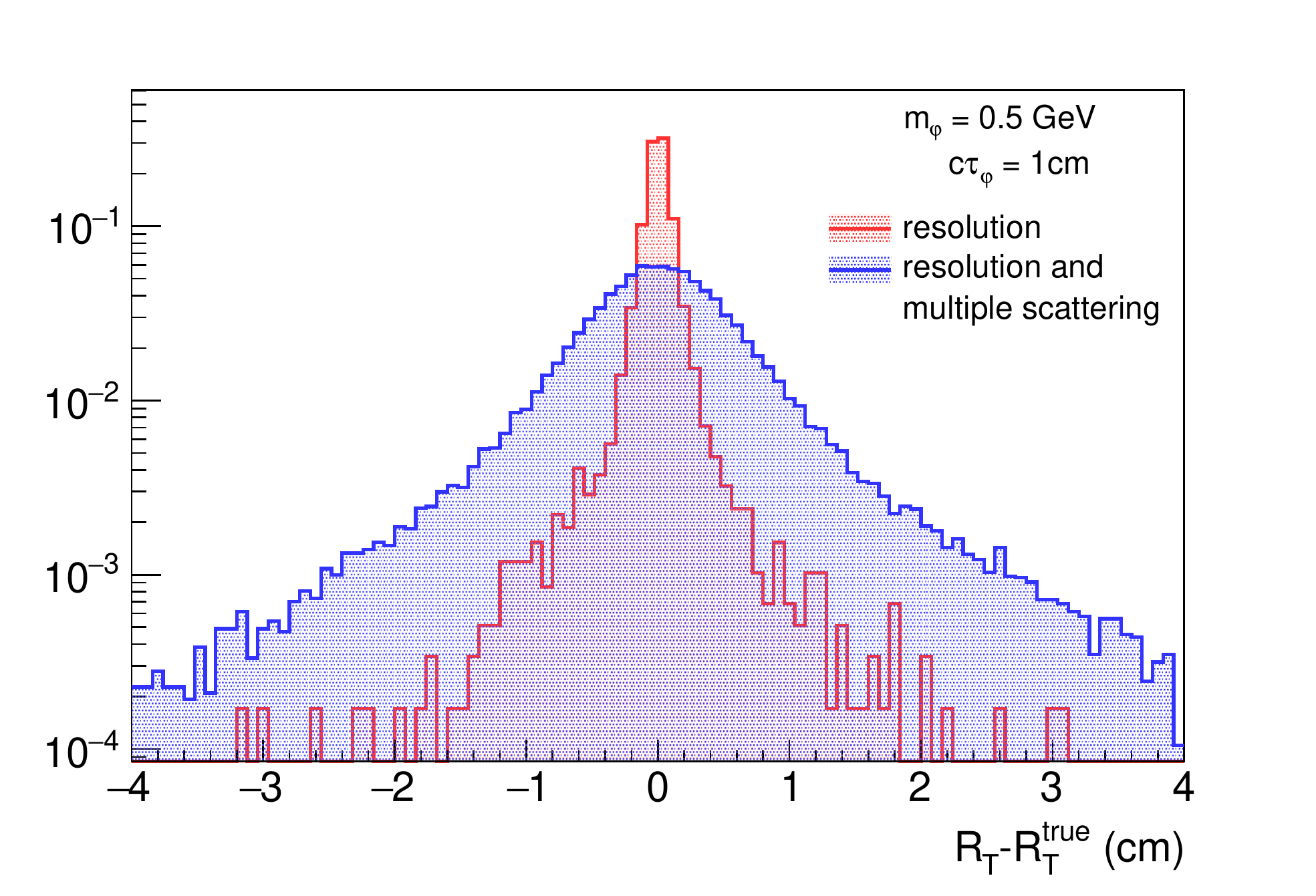}%
\caption{\label{fig:mcres_r} Effect of multiple scattering on vertex radius resolution, for a signal sample with $m_\varphi = 0.5$ GeV and \mbox{$c\tau_\varphi=1$ cm.} For this signal sample, the median $p_T$ of $\phi$ is 5.8 GeV}
\end{figure}

Once a candidate vertex is found, we use its coordinates and the reconstructed track momenta to compute the impact parameter of the mother particle's trajectory in the transverse plane with respect to the origin ($d_T$). We further compute the angle $(\alpha_T)$ in the transverse plane between the mother's reconstructed trajectory and the line connecting the vertex location with the origin. Finally, we record the distance of the vertex from the origin, in the transverse plane ($R_T$). The above variables are summarized in the diagram in Fig.~\ref{fig:vertex_params}. Note that we use the origin of the CMS coordinate system as a reference point, rather than the primary vertex, since we do not assume we identified the primary vertex at the trigger level. Fig.~\ref{fig:mcres_r} shows the resolution of the reconstructed vertex radius, with and without multiple scattering, for $m_\varphi = 0.5$ GeV.

\section{\label{sec:analysis}Analysis and Results}
\begin{figure}[b]
\includegraphics[width=0.4\textwidth]{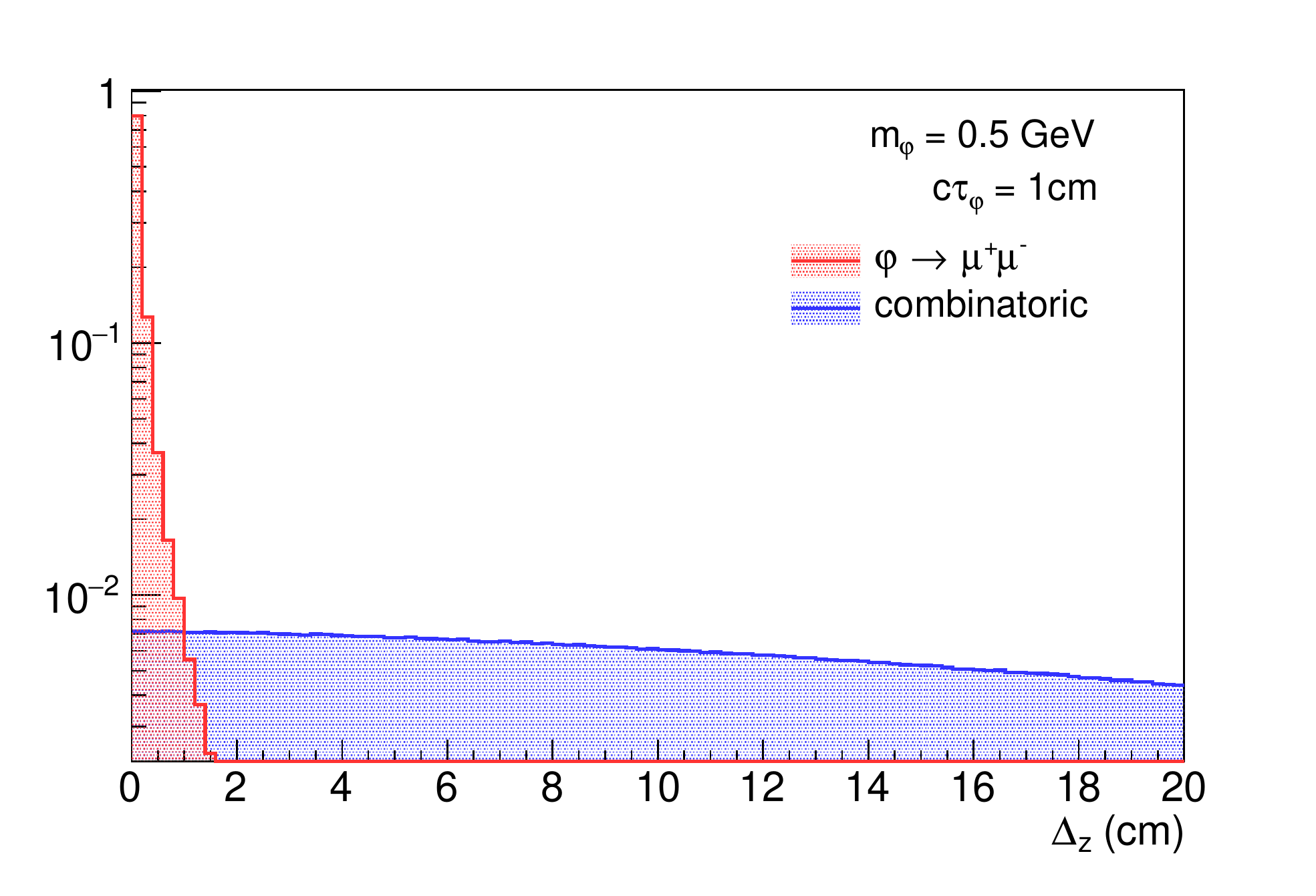}%
\caption{\label{fig:an_dz} The distance along the $z$ direction between the two helices at their intersection in the transverse plane, for a signal with $m_\varphi=0.5$ GeV and $c\tau=1$ cm (red) and random crossings by fake tracks (blue), after cuts i) through iii).}
\end{figure}

\begin{figure}[t]
\includegraphics[width=0.4\textwidth]{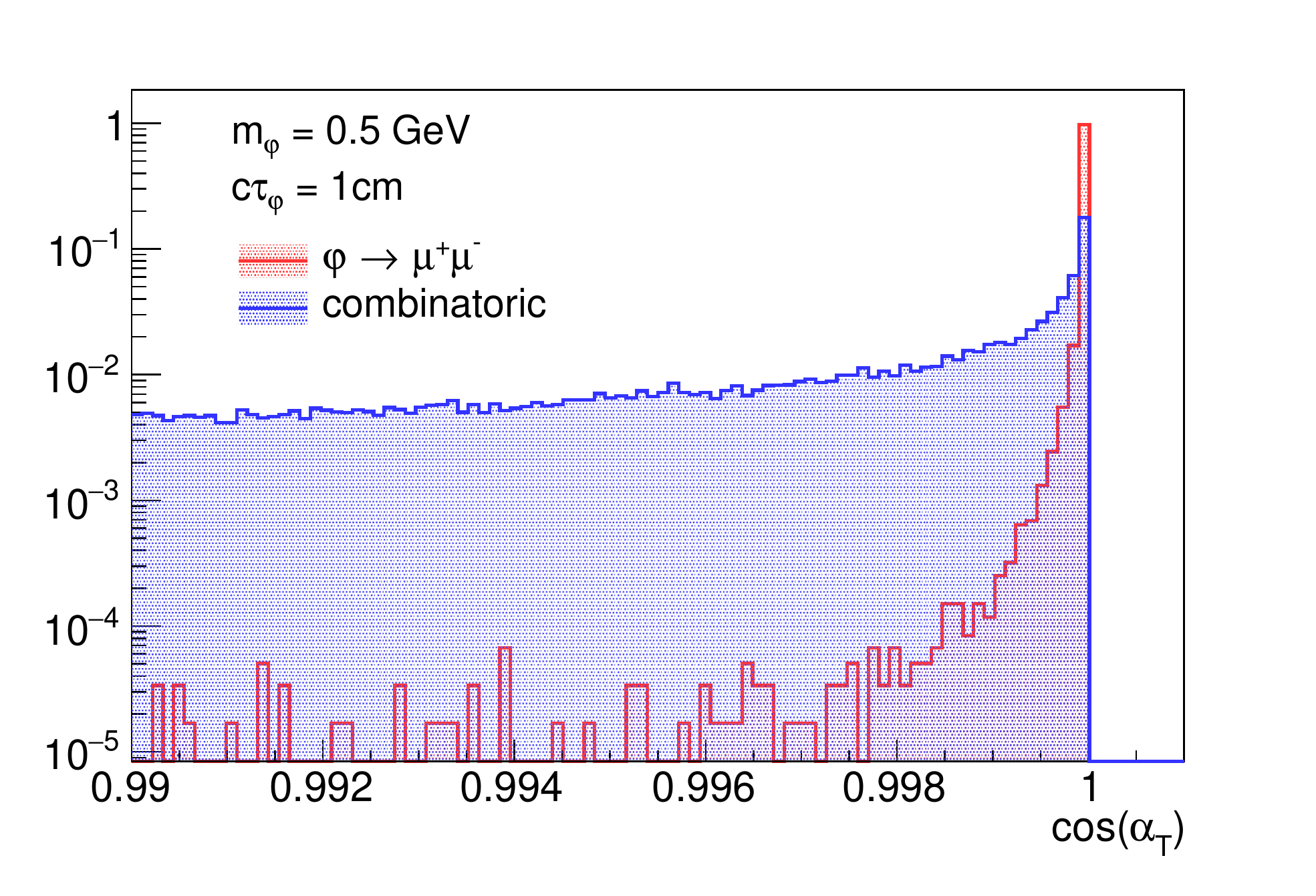}%
\caption{\label{fig:an_cosa} Cosine of the angle between $\varphi$'s reconstructed momentum and the line connecting the vertex location with the origin, in the transverse plane, for a signal with $m_\varphi=0.5$ GeV and $c\tau=1$ cm (red) and random crossings by fake tracks (blue), after cuts i) through iv).}
\end{figure}

\begin{figure}[t]
\includegraphics[width=0.4\textwidth]{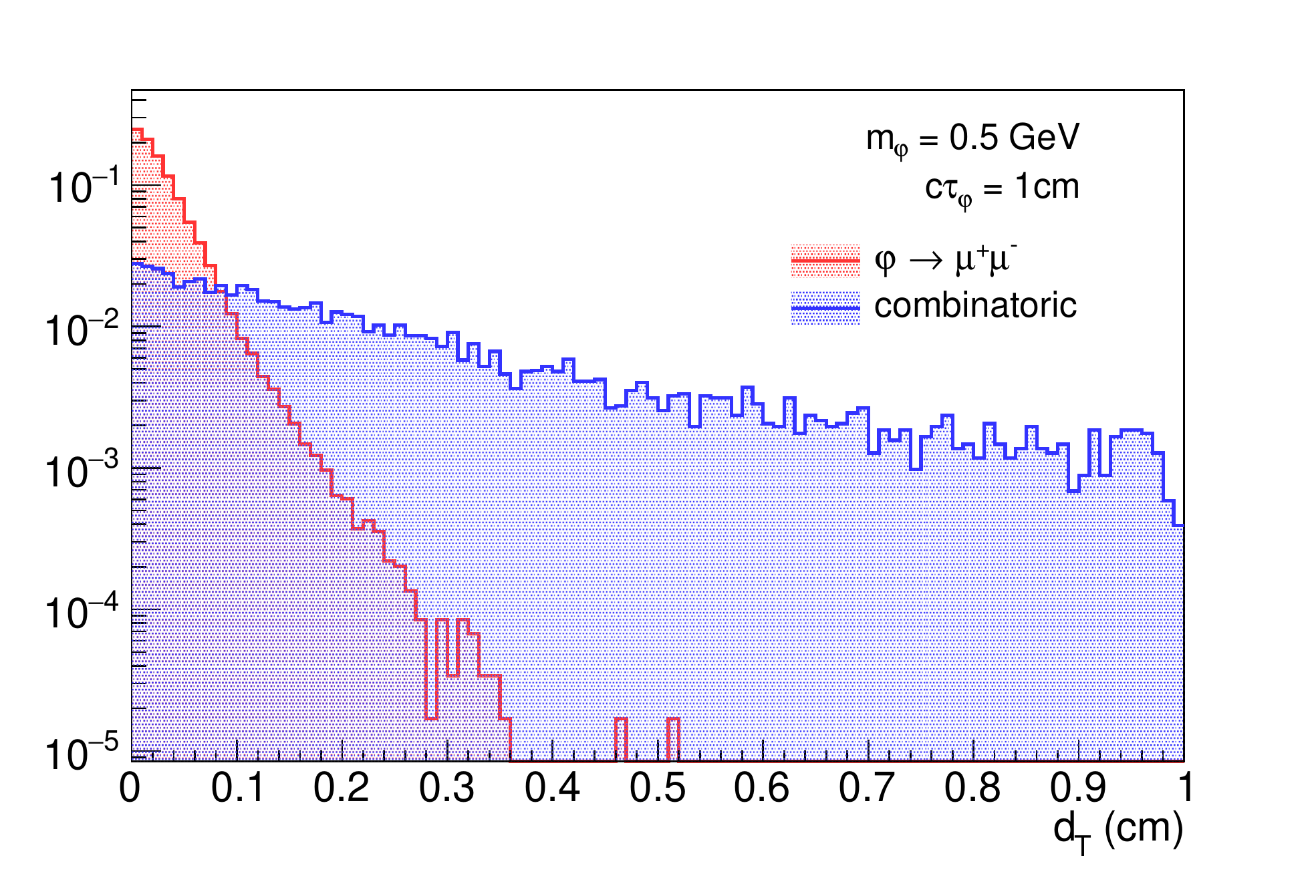}%
\caption{\label{fig:an_ipx} Impact parameter of the LLP candidate in the transverse plane, for a signal with $m_\varphi=0.5$ GeV and $c\tau=1$ cm (red) and random crossings by fake tracks (blue), after cuts i) through v).}
\end{figure}

Figs.~\ref{fig:an_dz}, \ref{fig:an_cosa} and \ref{fig:an_ipx} show the distributions of $\Delta_z$, $\cos\alpha_T$ and $d_T$ respectively, for the signal and for fake vertices. Based on these distributions, we define the following set of cuts:
\begin{enumerate}
\itemsep-0.2em 
\item $d_0>1$ mm for each track
\item at least 5 stubs for one of the tracks, and at least 4 stubs for the other.
\item $\Delta_T<0.2$ cm
\item $\Delta_z<1.0$ cm
\item $ \cos\alpha_T<0.9995$ 
\item $ d_T<0.1$ cm
\item $ R_T>1.5$ cm.
\end{enumerate}
The cut on the track impact parameter $d_0$ in i) prevents prompt tracks from contributing to the rate. The transverse size of the beam spot is negligible compared to the $d_0$ resolution of the track trigger (see Fig.~\ref{fig:mcres_d0}). Cuts ii) to vi) are intended to suppress the contribution from fake tracks to the trigger rate. Of these iv) and v) provide roughly $10^{-2}$ background suppression each, as can be seen from Figs.~\ref{fig:an_dz} and \ref{fig:an_cosa}. On the other hand, the  background reduction from ii) and iii) combined is only $\sim 50\%$ and these cuts could in principle be omitted.  The cuts on $d_T$ and $\alpha_T$ are rather strongly correlated, but some additional background suppression is nevertheless gained by imposing both, with virtually no loss in signal efficiency (see Fig.~\ref{fig:an_ipx}).
 Finally, the cut on the vertex radius in vii) reduces the rate from B-decays to the kHz level, though it is to a large degree degenerate with the cuts on $d_0$. 
 
\begin{table}[b]
\caption{\label{tab:rates}
Rates for events to have at least one vertex candidate, before the muon system requirements. Pile up of 200 is assumed.}
\begin{ruledtabular}
\begin{tabular}{ccc}
 minimum $p_T$ selection & fakes & $K_S$\\
\hline
(2, 2) GeV & 0.050 & 0.114 \\
(3, 3) GeV & 0.025 & 0.020 \\
(4, 4) GeV & 0.015 & 0.006 \\
(5, 3) GeV & 0.021 & 0.005 \\
\end{tabular}
\end{ruledtabular}
\end{table}

\begin{figure*}[t]
\includegraphics[width=0.43\linewidth]{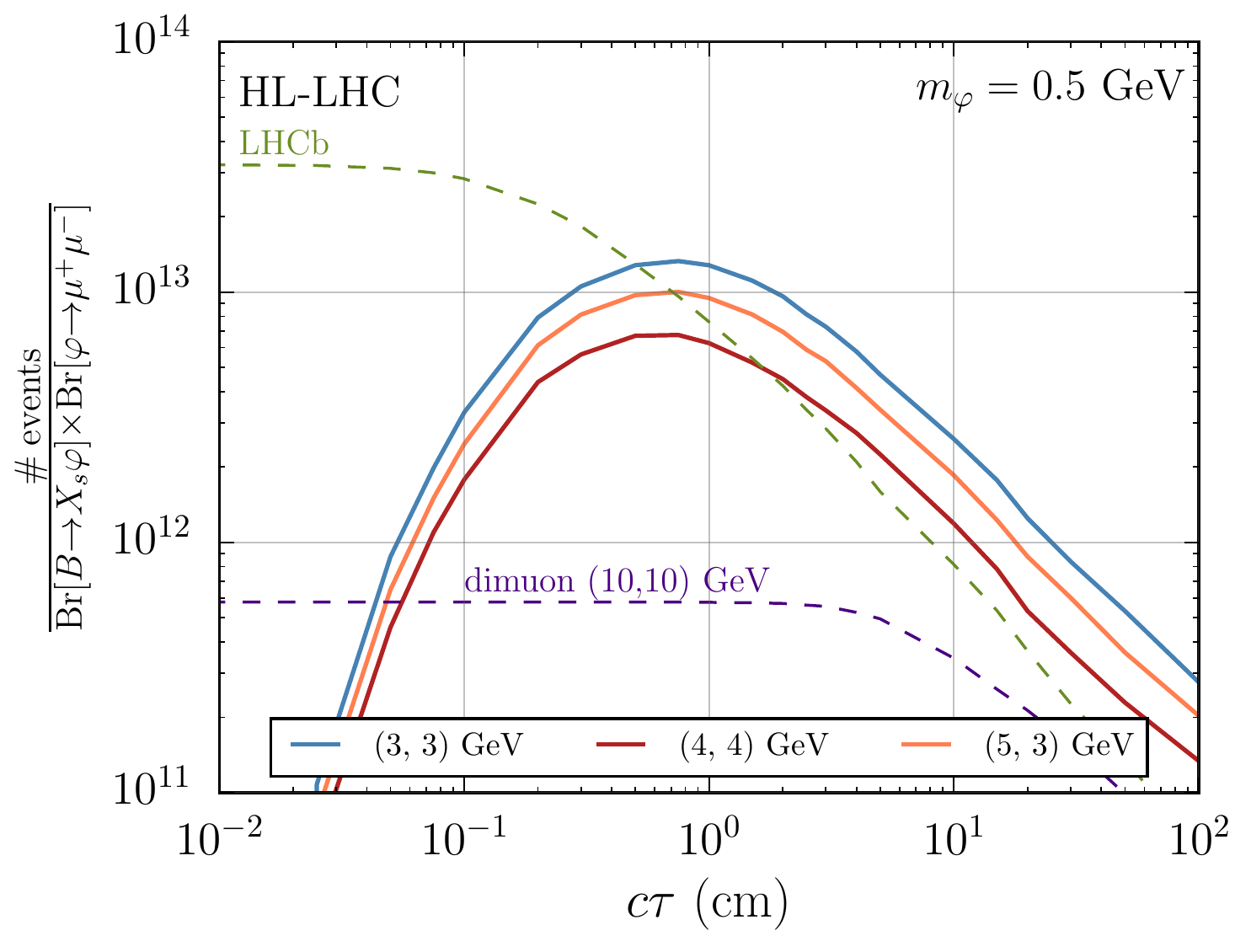}\hspace{1cm}
\includegraphics[width=0.43\linewidth]{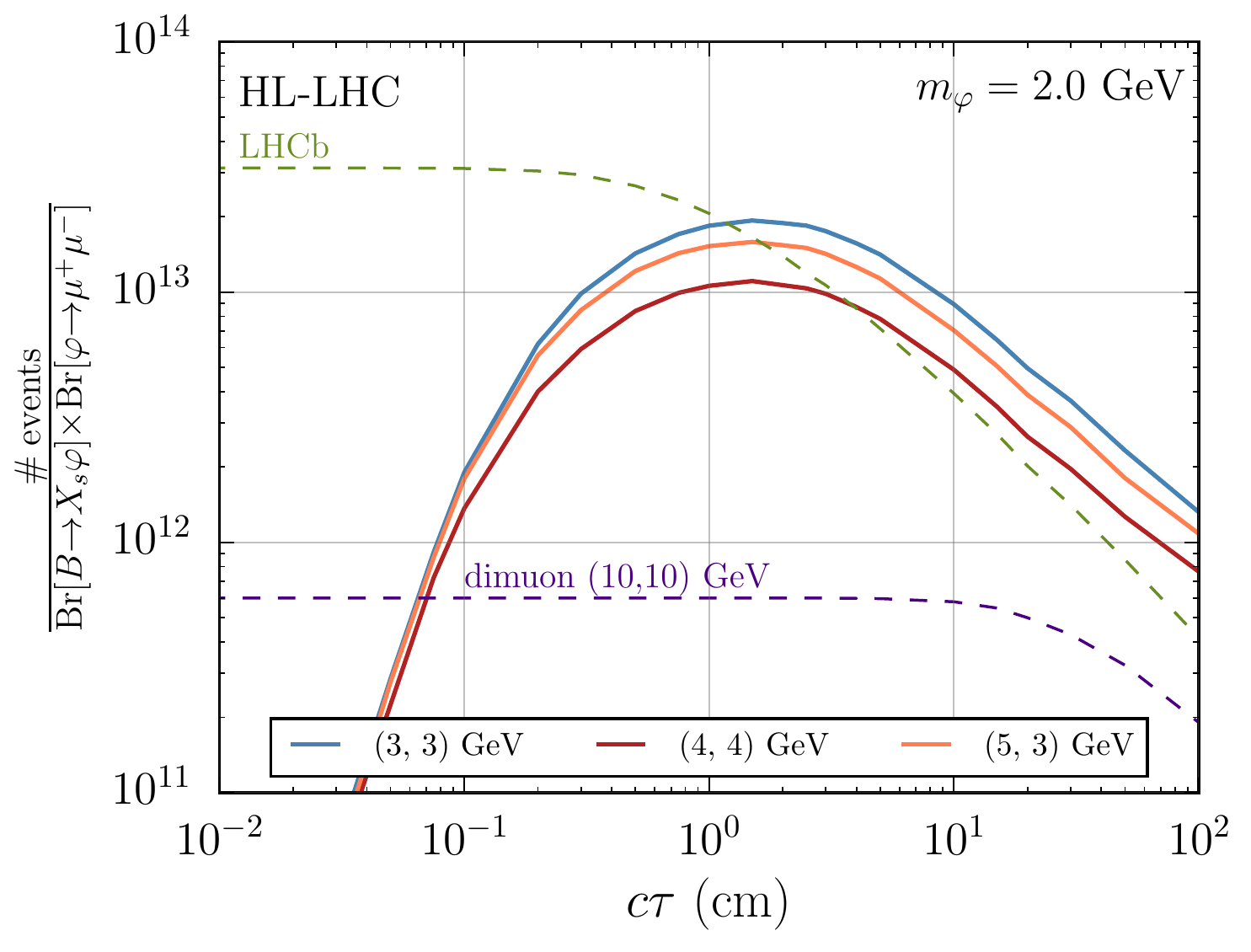}%
\caption{\label{fig:yield} Projected yield for the CMS displaced dimuon vertex trigger for three different choices of the muon $p_T$ threshold (solid), compared with LHCb (dashed green) and a hypothetical future ATLAS/CMS trigger on two standalone  muons with $p_T>$ 10 GeV each (dashed purple).}
\end{figure*}

The rates for finding a vertex satisfying the above criteria are given in Tab.~\ref{tab:rates} for various $p_T$ cuts on the tracks. Fake vertices and $K_S$ decay contribute roughly equally to the trigger rate.\footnote{Algorithms are being considered to identify the primary vertex already at the track trigger level. Given the relative softness of the $b$-$\bar b$ events, the primary vertex would still be misidentified a fraction of the time. If the associated loss in signal efficiency however proves to be acceptable, a cut on the longitudinal impact angle could be added (See Fig.~\ref{fig:alpha_l} in Appendix \ref{sec:turnon}). This would render the rate from fake vertices negligible compared to that from $K_S$ decays.}
As is, these rates are still too high for a realistic trigger and an additional selection is therefore needed to further reduce the rate with roughly two orders of magnitude. Depending on the signature of interest, this could for instance be an $H_T$ or MET requirement, or a second displaced vertex in the event. 

For our example, we require that both tracks are matched to a track candidate in the muon detectors. The signal efficiency for this requirement is difficult to model without a full simulation and depends on the $p_T$ threshold of the muons. We therefore assume that the signal efficiency is a step function with a 100\% efficient plateau, but vary the muon $p_T$-thresholds to account for this uncertainty. The background rate is driven by the fake muon rate: If the rate at which a fake track or $\pi^{\pm}$ mimics a muon is less that roughly 1/50 per track, the overall rate will be at the kHz level or lower. We consider this estimate for the muon fake rate conservative \cite{Collaboration:2283189}, though a full simulation with the CMS detector simulation is needed to validate it. It is also worth noting that the vertex requirement will greatly reduce the combinatorics for the muon detector matching, since only the tracks associated with a displaced vertex need to be checked.


Both ATLAS and CMS are aiming to include a standalone dimuon trigger in their HL-LHC trigger menus, with $p_T\gtrsim10$ GeV thresholds for both muons \cite{Collaboration:2283189,ATL-PHYS-PUB-2019-002}. As a point of comparison, we loosely model the acceptance of such a trigger by
\begin{enumerate}
\itemsep-0.2em 
\item $|\eta|<2.4$ for both muons
\item $p_T>10$ GeV  for both muons
\item $ R_T<600$ cm
\end{enumerate}
where the $R_T$ cut is loosely based on geometry of the ATLAS muon system. We assume that this trigger is 100\% efficient, which is conservative for the sake of our comparison. We further compare with the acceptance of LHCb, following the discussion in \cite{Ilten:2015hya}: 
\begin{enumerate}
\itemsep-0.2em 
\item $2.0<\eta<5$ for both tracks
\item $p_T>0.5$ GeV  for both tracks
\item 3-momentum of both tracks larger than 6 GeV
\item $ R_T<2.2$ cm
\item $z$-coordinate of vertex smaller 60 cm
\end{enumerate}
The cuts on the vertex location are motivated by the geometry of the VELO detector, however for the \mbox{$B\to X_s \varphi$} signature LHCb may reconstruct the $\varphi$ also somewhat outside the VELO by triggering on the associated $X_s$ state. This can increase the LHCb efficiency with an $\mathcal{O}(1)$ amount in the regime where the $\varphi$ predominantly decays outside the VELO \cite{williams}. We assumed a total integrated luminosity of $300\,\text{fb}^{-1}$ for LHCb and $3\,\text{ab}^{-1}$ for ATLAS and CMS.
 
The resulting yield for all three strategies is shown in Fig.~\ref{fig:yield} for two benchmark mass points, shown in terms of the production rate and lifetime of $\varphi$. Given that the suppression factor from the matching with the muon system is unknown at this time, we show the results for the CMS displaced vertex trigger for a few different $p_T$ thresholds on the muons. The (2,2) point in Tab.~\ref{tab:rates} was hereby omitted, since muons this soft will not be reconstructed by the muon chambers. With the assumptions above, we find that CMS has the potential to improve on the LHCb yield in the large $c\tau$ regime. In the low $c\tau$ regime LHCb clearly performs better due to the better vertex resolution of the VELO detector and LHCb's capability to do the full event reconstruction online. 

Our proposed displaced vertex trigger is moreover highly complementary to the standalone dimuon trigger (Dashed purple curve in Fig.~\ref{fig:yield}): At high $c\tau$ the standalone dimuon trigger has much larger geometric acceptance than our displaced vertex trigger, however for soft signals, such as heavy flavor decays, this is offset by our lower $p_T$ requirements. For high energy signals from e.g. supersymmetric particles, the standalone dimuon trigger will continue to have the best efficiency. 

It should be noted that all curves are each somewhat optimistic in different ways: For LHCb and the standalone dimuon trigger no detector smearing is attempted and the reconstruction is assumed to be 100\% efficient. For the displaced vertex trigger we have assumed negligible losses in signal efficiency at the high level trigger. Given that the backgrounds in all cases are highly non-trivial and are best estimated from data, we do not attempt to project exclusion limits at this stage.

\section{\label{sec:summary}Discussion}
We have performed a toy simulation of the future CMS track trigger for the heavy flavor benchmark $B\to X_s\varphi$, where $\varphi$ is a long-lived particle with an appreciable branching ratio to muons. The rate from random crossings of fake tracks and $K_S$ mesons can be brought under control by a combination of pointing cuts and by matching the tracks to activity in the muon chambers. The track trigger could allow CMS to compete with LHCb on soft, displaced signatures beyond the Standard Model physics, in particular in the context of exotic heavy flavor decays and dark photon models.

As mentioned above, at least two handles (pointing and muon matching) are needed to bring the rate from random track crossings down to an acceptable level for the high level trigger. There are therefore a number of possible variations on our strategy, by relaxing the pointing and/or muon requirements and instead demanding: 
\begin{itemize}
\item A dimuon vertex without pointing but with a moderate amount of MET. This may be powerful for inelastic dark matter models.
\item Two reconstructed vertices in the event. This is of interest for e.g.~hidden valley/dark shower models or models where the dark sector states are pair produced.
\item A vertex with 4 or more tracks, which would capture hadronic decays or decays to $\tau^+\tau^-$.  
\item A single, non-pointing vertex with one muon with moderate $p_T$, which can record decays of heavy neutral leptons. 
\end{itemize}
We leave these and other possibilities for future work.

\begin{acknowledgments}
We are grateful to Rebecca Carney, Yangyang Cheng and Simone Pagan Griso for collaboration in the early phase of the project and for useful discussions afterwards. We also thank Michalis Bachtis, David Curtin, Vladimir V. Gligorov, Sergo Jindariani, Zhen Liu, Jacobo Konigsberg, Jessie Shelton, Scott Thomas, Dong Xu and the participants of the ``New ideas in detecting long-lived particles at the LHC'' workshop (LBNL, July 2018) for comments and discussions, and Vladimir V. Gligorov, Zhen Liu, Keith Ulmer and Michael Williams for comments on the manuscript. YG is supported by NSF grant number 1913356 and SK is supported by DOE grant DE-SC0009988. This research used resources of the National Energy Research Scientific Computing Center (NERSC), a U.S. Department of Energy Office of Science User Facility operated under Contract No. DE-AC02-05CH11231.

\end{acknowledgments}

\bibliography{apssamp}

\providecommand{\noopsort}[1]{}\providecommand{\singleletter}[1]{#1}%
\begin{thebibliography}{19}%
\makeatletter
\providecommand \@ifxundefined [1]{%
 \@ifx{#1\undefined}
}%
\providecommand \@ifnum [1]{%
 \ifnum #1\expandafter \@firstoftwo
 \else \expandafter \@secondoftwo
 \fi
}%
\providecommand \@ifx [1]{%
 \ifx #1\expandafter \@firstoftwo
 \else \expandafter \@secondoftwo
 \fi
}%
\providecommand \natexlab [1]{#1}%
\providecommand \enquote  [1]{``#1''}%
\providecommand \bibnamefont  [1]{#1}%
\providecommand \bibfnamefont [1]{#1}%
\providecommand \citenamefont [1]{#1}%
\providecommand \href@noop [0]{\@secondoftwo}%
\providecommand \href [0]{\begingroup \@sanitize@url \@href}%
\providecommand \@href[1]{\@@startlink{#1}\@@href}%
\providecommand \@@href[1]{\endgroup#1\@@endlink}%
\providecommand \@sanitize@url [0]{\catcode `\\12\catcode `\$12\catcode
  `\&12\catcode `\#12\catcode `\^12\catcode `\_12\catcode `\%12\relax}%
\providecommand \@@startlink[1]{}%
\providecommand \@@endlink[0]{}%
\providecommand \url  [0]{\begingroup\@sanitize@url \@url }%
\providecommand \@url [1]{\endgroup\@href {#1}{\urlprefix }}%
\providecommand \urlprefix  [0]{URL }%
\providecommand \Eprint [0]{\href }%
\providecommand \doibase [0]{https://doi.org/}%
\providecommand \selectlanguage [0]{\@gobble}%
\providecommand \bibinfo  [0]{\@secondoftwo}%
\providecommand \bibfield  [0]{\@secondoftwo}%
\providecommand \translation [1]{[#1]}%
\providecommand \BibitemOpen [0]{}%
\providecommand \bibitemStop [0]{}%
\providecommand \bibitemNoStop [0]{.\EOS\space}%
\providecommand \EOS [0]{\spacefactor3000\relax}%
\providecommand \BibitemShut  [1]{\csname bibitem#1\endcsname}%
\let\auto@bib@innerbib\@empty
\bibitem [{\citenamefont {Einsweiler}\ \emph {et~al.}(2017)\citenamefont
  {Einsweiler} \emph {et~al.}}]{Collaboration:2285585}%
  \BibitemOpen
  \bibfield  {author} {\bibinfo {author} {\bibfnamefont {K.}~\bibnamefont
  {Einsweiler}} \emph {et~al.} (\bibinfo {collaboration} {ATLAS
  collaboration}),\ }\href {https://cds.cern.ch/record/2285585} {\emph
  {\bibinfo {title} {{Technical Design Report for the ATLAS Inner Tracker Pixel
  Detector}}}},\ \bibinfo {type} {Tech. Rep.}\ \bibinfo {number}
  {CERN-LHCC-2017-021. ATLAS-TDR-030}\ (\bibinfo  {institution} {CERN},\
  \bibinfo {address} {Geneva},\ \bibinfo {year} {2017})\BibitemShut {NoStop}%
\bibitem [{\citenamefont {Contardo}\ \emph {et~al.}(2015)\citenamefont
  {Contardo}, \citenamefont {Klute}, \citenamefont {Mans}, \citenamefont
  {Silvestris},\ and\ \citenamefont {Butler}}]{CMSCollaboration:2015zni}%
  \BibitemOpen
  \bibfield  {author} {\bibinfo {author} {\bibfnamefont {D.}~\bibnamefont
  {Contardo}}, \bibinfo {author} {\bibfnamefont {M.}~\bibnamefont {Klute}},
  \bibinfo {author} {\bibfnamefont {J.}~\bibnamefont {Mans}}, \bibinfo {author}
  {\bibfnamefont {L.}~\bibnamefont {Silvestris}}, and\ \bibinfo {author}
  {\bibfnamefont {J.}~\bibnamefont {Butler}},\ }\href
  {https://cds.cern.ch/record/2020886?ln=en} {\emph {\bibinfo {title}
  {{Technical Proposal for the Phase-II Upgrade of the CMS Detector}}}},\
  \bibinfo {type} {Tech. Rep.}\ \bibinfo {number} {CMS-TDR-15-02}\ (\bibinfo
  {year} {2015})\BibitemShut {NoStop}%
\bibitem [{\citenamefont {Bediaga}\ \emph {et~al.}(2012)\citenamefont {Bediaga}
  \emph {et~al.}}]{Bediaga:1443882}%
  \BibitemOpen
  \bibfield  {author} {\bibinfo {author} {\bibfnamefont {I.}~\bibnamefont
  {Bediaga}} \emph {et~al.} (\bibinfo {collaboration} {LHCb collaboration}),\
  }\href {https://cds.cern.ch/record/1443882} {\emph {\bibinfo {title}
  {{Framework TDR for the LHCb Upgrade: Technical Design Report}}}},\ \bibinfo
  {type} {Tech. Rep.}\ \bibinfo {number} {CERN-LHCC-2012-007. LHCb-TDR-12}\
  (\bibinfo {year} {2012})\BibitemShut {NoStop}%
\bibitem [{\citenamefont {Klein}\ \emph {et~al.}(2017)\citenamefont {Klein}
  \emph {et~al.}}]{Collaboration:2272264}%
  \BibitemOpen
  \bibfield  {author} {\bibinfo {author} {\bibfnamefont {K.}~\bibnamefont
  {Klein}} \emph {et~al.} (\bibinfo {collaboration} {CMS collaboration}),\
  }\href {https://cds.cern.ch/record/2272264} {\emph {\bibinfo {title} {{The
  Phase-2 Upgrade of the CMS Tracker}}}},\ \bibinfo {type} {Tech. Rep.}\
  \bibinfo {number} {CERN-LHCC-2017-009. CMS-TDR-014}\ (\bibinfo  {institution}
  {CERN},\ \bibinfo {address} {Geneva},\ \bibinfo {year} {2017})\BibitemShut
  {NoStop}%
\bibitem [{\citenamefont {Gershtein}(2017)}]{Gershtein:2017tsv}%
  \BibitemOpen
  \bibfield  {author} {\bibinfo {author} {\bibfnamefont {Y.}~\bibnamefont
  {Gershtein}},\ }\bibfield  {title} {\bibinfo {title} {{CMS Hardware Track
  Trigger: New Opportunities for Long-Lived Particle Searches at the HL-LHC}},\
  }\href {https://doi.org/10.1103/PhysRevD.96.035027} {\bibfield  {journal}
  {\bibinfo  {journal} {Phys. Rev.}\ }\textbf {\bibinfo {volume} {D96}},\
  \bibinfo {pages} {035027} (\bibinfo {year} {2017})},\ \Eprint
  {https://arxiv.org/abs/1705.04321} {arXiv:1705.04321 [hep-ph]} \BibitemShut
  {NoStop}%
\bibitem [{\citenamefont {{CMS Collaboration}}(2018)}]{CMS-PAS-FTR-18-018}%
  \BibitemOpen
  \bibfield  {author} {\bibinfo {author} {\bibnamefont {{CMS Collaboration}}},\
  }\href {https://cds.cern.ch/record/2647987} {\emph {\bibinfo {title} {{First
  Level Track Jet Trigger for Displaced Jets at High Luminosity LHC}}}},\
  \bibinfo {type} {Tech. Rep.}\ \bibinfo {number} {CMS-PAS-FTR-18-018}\
  (\bibinfo {address} {Geneva},\ \bibinfo {year} {2018})\BibitemShut {NoStop}%
\bibitem [{\citenamefont {Aaij}\ \emph
  {et~al.}(2017{\natexlab{a}})\citenamefont {Aaij} \emph
  {et~al.}}]{Aaij:2016qsm}%
  \BibitemOpen
  \bibfield  {author} {\bibinfo {author} {\bibfnamefont {R.}~\bibnamefont
  {Aaij}} \emph {et~al.} (\bibinfo {collaboration} {LHCb collaboration}),\
  }\bibfield  {title} {\bibinfo {title} {{Search for long-lived scalar
  particles in $B^+ \to K^+ \chi (\mu^+\mu^-)$ decays}},\ }\href
  {https://doi.org/10.1103/PhysRevD.95.071101} {\bibfield  {journal} {\bibinfo
  {journal} {Phys. Rev.}\ }\textbf {\bibinfo {volume} {D95}},\ \bibinfo {pages}
  {071101} (\bibinfo {year} {2017}{\natexlab{a}})},\ \Eprint
  {https://arxiv.org/abs/1612.07818} {arXiv:1612.07818 [hep-ex]} \BibitemShut
  {NoStop}%
\bibitem [{\citenamefont {Aaij}\ \emph {et~al.}(2015)\citenamefont {Aaij} \emph
  {et~al.}}]{Aaij:2015tna}%
  \BibitemOpen
  \bibfield  {author} {\bibinfo {author} {\bibfnamefont {R.}~\bibnamefont
  {Aaij}} \emph {et~al.} (\bibinfo {collaboration} {LHCb collaboration}),\
  }\bibfield  {title} {\bibinfo {title} {{Search for hidden-sector bosons in
  $B^0 \!\to K^{*0}\mu^+\mu^-$ decays}},\ }\href
  {https://doi.org/10.1103/PhysRevLett.115.161802} {\bibfield  {journal}
  {\bibinfo  {journal} {Phys. Rev. Lett.}\ }\textbf {\bibinfo {volume} {115}},\
  \bibinfo {pages} {161802} (\bibinfo {year} {2015})},\ \Eprint
  {https://arxiv.org/abs/1508.04094} {arXiv:1508.04094 [hep-ex]} \BibitemShut
  {NoStop}%
\bibitem [{\citenamefont {Aaij}\ \emph {et~al.}(2018)\citenamefont {Aaij} \emph
  {et~al.}}]{Aaij:2017rft}%
  \BibitemOpen
  \bibfield  {author} {\bibinfo {author} {\bibfnamefont {R.}~\bibnamefont
  {Aaij}} \emph {et~al.} (\bibinfo {collaboration} {LHCb collaboration}),\
  }\bibfield  {title} {\bibinfo {title} {{Search for Dark Photons Produced in
  13 TeV $pp$ Collisions}},\ }\href
  {https://doi.org/10.1103/PhysRevLett.120.061801} {\bibfield  {journal}
  {\bibinfo  {journal} {Phys. Rev. Lett.}\ }\textbf {\bibinfo {volume} {120}},\
  \bibinfo {pages} {061801} (\bibinfo {year} {2018})},\ \Eprint
  {https://arxiv.org/abs/1710.02867} {arXiv:1710.02867 [hep-ex]} \BibitemShut
  {NoStop}%
\bibitem [{\citenamefont {Ilten}\ \emph {et~al.}(2016)\citenamefont {Ilten},
  \citenamefont {Soreq}, \citenamefont {Thaler}, \citenamefont {Williams},\
  and\ \citenamefont {Xue}}]{Ilten:2016tkc}%
  \BibitemOpen
  \bibfield  {author} {\bibinfo {author} {\bibfnamefont {P.}~\bibnamefont
  {Ilten}}, \bibinfo {author} {\bibfnamefont {Y.}~\bibnamefont {Soreq}},
  \bibinfo {author} {\bibfnamefont {J.}~\bibnamefont {Thaler}}, \bibinfo
  {author} {\bibfnamefont {M.}~\bibnamefont {Williams}}, and\ \bibinfo {author}
  {\bibfnamefont {W.}~\bibnamefont {Xue}},\ }\bibfield  {title} {\bibinfo
  {title} {{Proposed Inclusive Dark Photon Search at LHCb}},\ }\href
  {https://doi.org/10.1103/PhysRevLett.116.251803} {\bibfield  {journal}
  {\bibinfo  {journal} {Phys. Rev. Lett.}\ }\textbf {\bibinfo {volume} {116}},\
  \bibinfo {pages} {251803} (\bibinfo {year} {2016})},\ \Eprint
  {https://arxiv.org/abs/1603.08926} {arXiv:1603.08926 [hep-ph]} \BibitemShut
  {NoStop}%
\bibitem [{\citenamefont {Ilten}\ \emph {et~al.}(2015)\citenamefont {Ilten},
  \citenamefont {Thaler}, \citenamefont {Williams},\ and\ \citenamefont
  {Xue}}]{Ilten:2015hya}%
  \BibitemOpen
  \bibfield  {author} {\bibinfo {author} {\bibfnamefont {P.}~\bibnamefont
  {Ilten}}, \bibinfo {author} {\bibfnamefont {J.}~\bibnamefont {Thaler}},
  \bibinfo {author} {\bibfnamefont {M.}~\bibnamefont {Williams}}, and\ \bibinfo
  {author} {\bibfnamefont {W.}~\bibnamefont {Xue}},\ }\bibfield  {title}
  {\bibinfo {title} {{Dark photons from charm mesons at LHCb}},\ }\href
  {https://doi.org/10.1103/PhysRevD.92.115017} {\bibfield  {journal} {\bibinfo
  {journal} {Phys. Rev.}\ }\textbf {\bibinfo {volume} {D92}},\ \bibinfo {pages}
  {115017} (\bibinfo {year} {2015})},\ \Eprint
  {https://arxiv.org/abs/1509.06765} {arXiv:1509.06765 [hep-ph]} \BibitemShut
  {NoStop}%
\bibitem [{\citenamefont {Sjostrand}\ \emph {et~al.}(2006)\citenamefont
  {Sjostrand}, \citenamefont {Mrenna},\ and\ \citenamefont
  {Skands}}]{Sjostrand:2006za}%
  \BibitemOpen
  \bibfield  {author} {\bibinfo {author} {\bibfnamefont {T.}~\bibnamefont
  {Sjostrand}}, \bibinfo {author} {\bibfnamefont {S.}~\bibnamefont {Mrenna}},
  and\ \bibinfo {author} {\bibfnamefont {P.~Z.}\ \bibnamefont {Skands}},\
  }\bibfield  {title} {\bibinfo {title} {{PYTHIA 6.4 Physics and Manual}},\
  }\href {https://doi.org/10.1088/1126-6708/2006/05/026} {\bibfield  {journal}
  {\bibinfo  {journal} {JHEP}\ }\textbf {\bibinfo {volume} {05}},\ \bibinfo
  {pages} {026}},\ \Eprint {https://arxiv.org/abs/hep-ph/0603175}
  {arXiv:hep-ph/0603175 [hep-ph]} \BibitemShut {NoStop}%
\bibitem [{\citenamefont {Sjostrand}\ \emph {et~al.}(2015)\citenamefont
  {Sjostrand} \emph {et~al.}}]{Sjostrand:2014zea}%
  \BibitemOpen
  \bibfield  {author} {\bibinfo {author} {\bibfnamefont {T.}~\bibnamefont
  {Sjostrand}} \emph {et~al.},\ }\bibfield  {title} {\bibinfo {title} {{An
  Introduction to PYTHIA 8.2}},\ }\href
  {https://doi.org/10.1016/j.cpc.2015.01.024} {\bibfield  {journal} {\bibinfo
  {journal} {Comput. Phys. Commun.}\ }\textbf {\bibinfo {volume} {191}},\
  \bibinfo {pages} {159} (\bibinfo {year} {2015})},\ \Eprint
  {https://arxiv.org/abs/1410.3012} {arXiv:1410.3012 [hep-ph]} \BibitemShut
  {NoStop}%
\bibitem [{\citenamefont {Aaij}\ \emph
  {et~al.}(2017{\natexlab{b}})\citenamefont {Aaij} \emph
  {et~al.}}]{Aaij:2016avz}%
  \BibitemOpen
  \bibfield  {author} {\bibinfo {author} {\bibfnamefont {R.}~\bibnamefont
  {Aaij}} \emph {et~al.} (\bibinfo {collaboration} {LHCb collaboration}),\
  }\bibfield  {title} {\bibinfo {title} {{Measurement of the $b$-quark
  production cross-section in 7 and 13 TeV $pp$ collisions}},\ }\href
  {https://doi.org/10.1103/PhysRevLett.119.169901,
  10.1103/PhysRevLett.118.052002} {\bibfield  {journal} {\bibinfo  {journal}
  {Phys. Rev. Lett.}\ }\textbf {\bibinfo {volume} {118}},\ \bibinfo {pages}
  {052002} (\bibinfo {year} {2017}{\natexlab{b}})},\ \bibinfo {note} {[Erratum:
  Phys. Rev. Lett.119,no.16,169901(2017)]},\ \Eprint
  {https://arxiv.org/abs/1612.05140} {arXiv:1612.05140 [hep-ex]} \BibitemShut
  {NoStop}%
\bibitem [{\citenamefont {Drevermann}(2007)}]{trackparam}%
  \BibitemOpen
  \bibfield  {author} {\bibinfo {author} {\bibfnamefont {H.}~\bibnamefont
  {Drevermann}},\ }\href
  {http://www.hep.ucl.ac.uk/atlas/atlantis/files/helix_equations_1.pdf}
  {\bibinfo {title} {Helix equations}} (\bibinfo {year} {April
  2007})\BibitemShut {NoStop}%
\bibitem [{\citenamefont {Tanabashi}\ \emph {et~al.}(2018)\citenamefont
  {Tanabashi} \emph {et~al.}}]{Tanabashi:2018oca}%
  \BibitemOpen
  \bibfield  {author} {\bibinfo {author} {\bibfnamefont {M.}~\bibnamefont
  {Tanabashi}} \emph {et~al.} (\bibinfo {collaboration} {Particle Data
  Group}),\ }\bibfield  {title} {\bibinfo {title} {{Review of Particle Physics,
  section 33.3}},\ }\href {https://doi.org/10.1103/PhysRevD.98.030001}
  {\bibfield  {journal} {\bibinfo  {journal} {Phys. Rev.}\ }\textbf {\bibinfo
  {volume} {D98}},\ \bibinfo {pages} {030001} (\bibinfo {year}
  {2018})}\BibitemShut {NoStop}%
\bibitem [{\citenamefont {Hebbeker}\ \emph {et~al.}(2017)\citenamefont
  {Hebbeker} \emph {et~al.}}]{Collaboration:2283189}%
  \BibitemOpen
  \bibfield  {author} {\bibinfo {author} {\bibfnamefont {T.}~\bibnamefont
  {Hebbeker}} \emph {et~al.} (\bibinfo {collaboration} {CMS collaboration}),\
  }\href {https://cds.cern.ch/record/2283189} {\emph {\bibinfo {title} {{The
  Phase-2 Upgrade of the CMS Muon Detectors}}}},\ \bibinfo {type} {Tech. Rep.}\
  \bibinfo {number} {CERN-LHCC-2017-012. CMS-TDR-016}\ (\bibinfo  {institution}
  {CERN},\ \bibinfo {address} {Geneva},\ \bibinfo {year} {2017})\BibitemShut
  {NoStop}%
\bibitem [{\citenamefont {{ATLAS
  Collaboration}}(2019)}]{ATL-PHYS-PUB-2019-002}%
  \BibitemOpen
  \bibfield  {author} {\bibinfo {author} {\bibnamefont {{ATLAS
  Collaboration}}},\ }\href {https://cds.cern.ch/record/2654518} {\emph
  {\bibinfo {title} {{Search prospects for dark-photons decaying to displaced
  collimated jets of muons at HL-LHC}}}},\ \bibinfo {type} {Tech. Rep.}\
  \bibinfo {number} {ATL-PHYS-PUB-2019-002}\ (\bibinfo  {institution} {CERN},\
  \bibinfo {address} {Geneva},\ \bibinfo {year} {2019})\BibitemShut {NoStop}%
\bibitem [{\citenamefont {Williams}()}]{williams}%
  \BibitemOpen
  \bibfield  {author} {\bibinfo {author} {\bibfnamefont {M.}~\bibnamefont
  {Williams}},\ }\href@noop {} {\bibinfo {title} {Private
  communication}}\BibitemShut {NoStop}%
\end{thebibliography}%

\appendix

\section{Additional figures\label{sec:turnon}}

\begin{figure}[h!]
\vspace{0.2cm}
\includegraphics[width=0.45\textwidth]{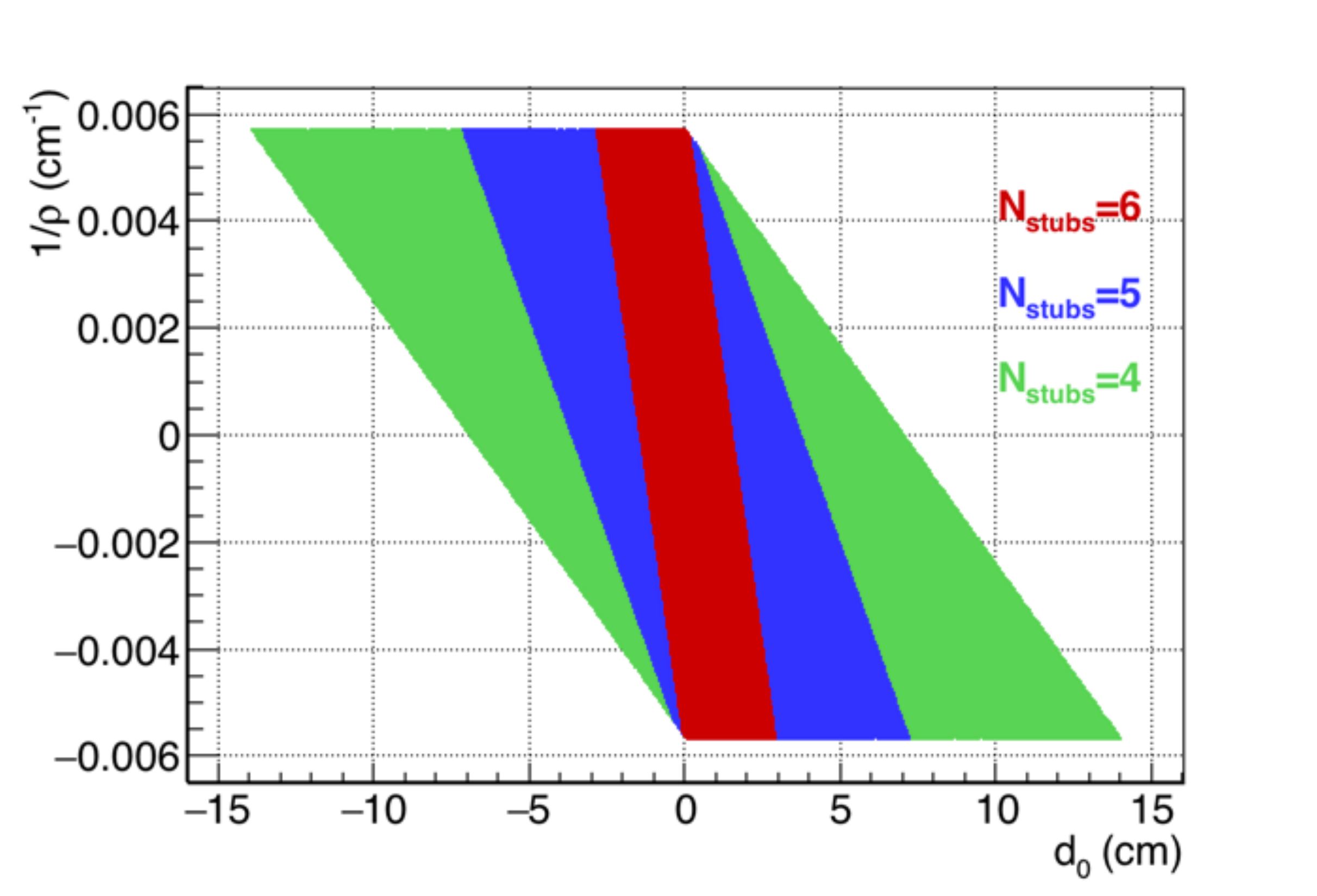}%
\caption{\label{fig:fake_pars} Expected number of stubs for a track given its curvature and impact parameter.}
\end{figure}

Fig.~\ref{fig:fake_pars} shows the correlation between the inverse track radius and the transverse impact parameter of the fake tracks, for tracks that are reconstructed.

\begin{figure}[h!]
\includegraphics[width=0.45\textwidth]{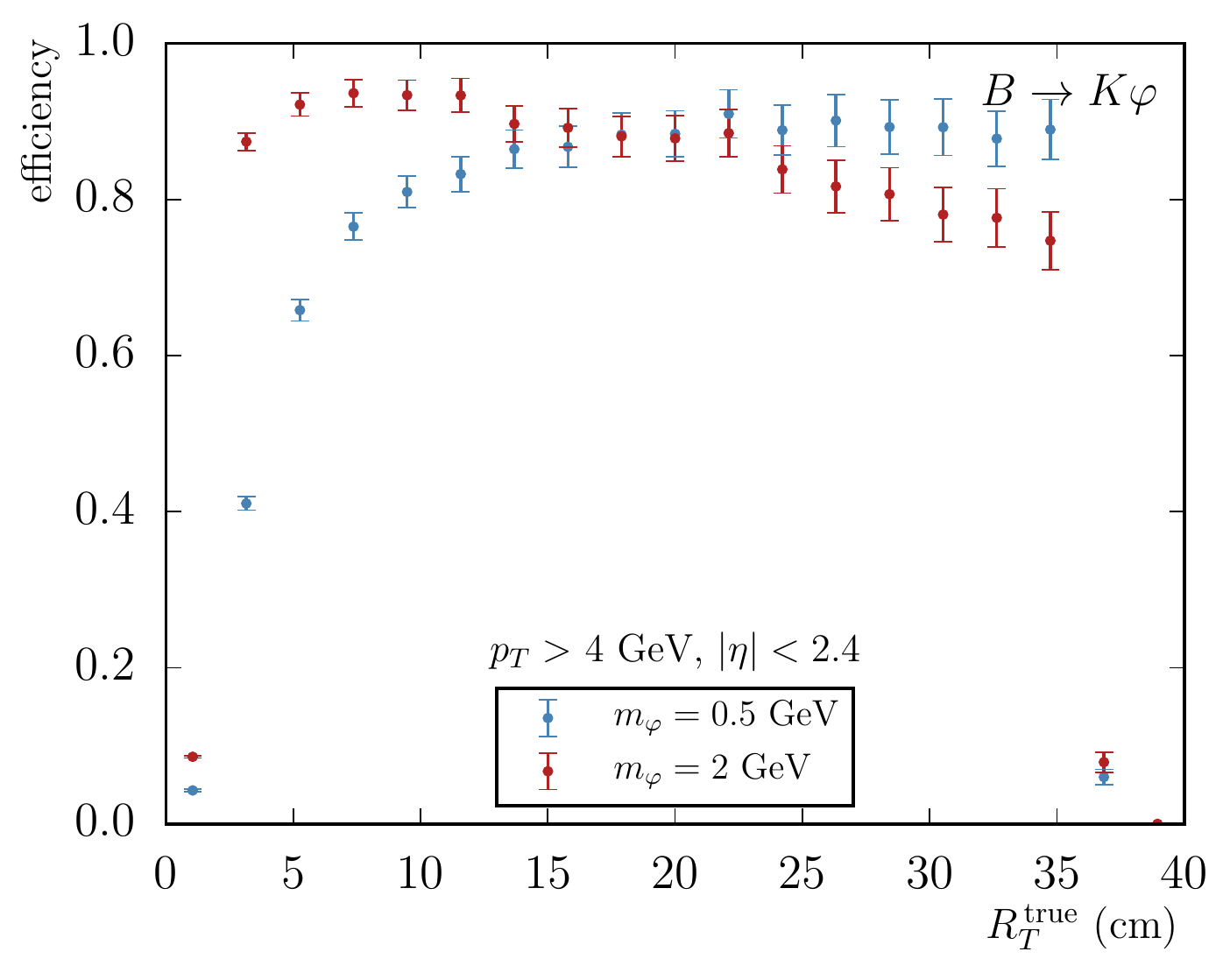}%
\caption{\label{fig:turn_on_RT} Trigger efficiency as a function of $R^{\text{true}}_T$, calculated after imposing fiducial cuts on both muons.}
\end{figure}

In Fig.~\ref{fig:turn_on_RT} we show the trigger efficiency as a function of the truth-level $R_T$, after imposing (truth-level) acceptance cuts on the $p_T$ and $\eta$ range for both muons. For higher masses and high $R_T$, the higher opening angle between the muons implies that the stubs are more likely to fail the cut on the pitch, such that insufficient stubs are found to reconstruct the track. At low radii, the $m_\varphi=0.5$ GeV benchmark point looses efficiency faster due to the $d_0$ cut on the tracks.

\begin{figure}[h!]
\includegraphics[width=0.45\textwidth]{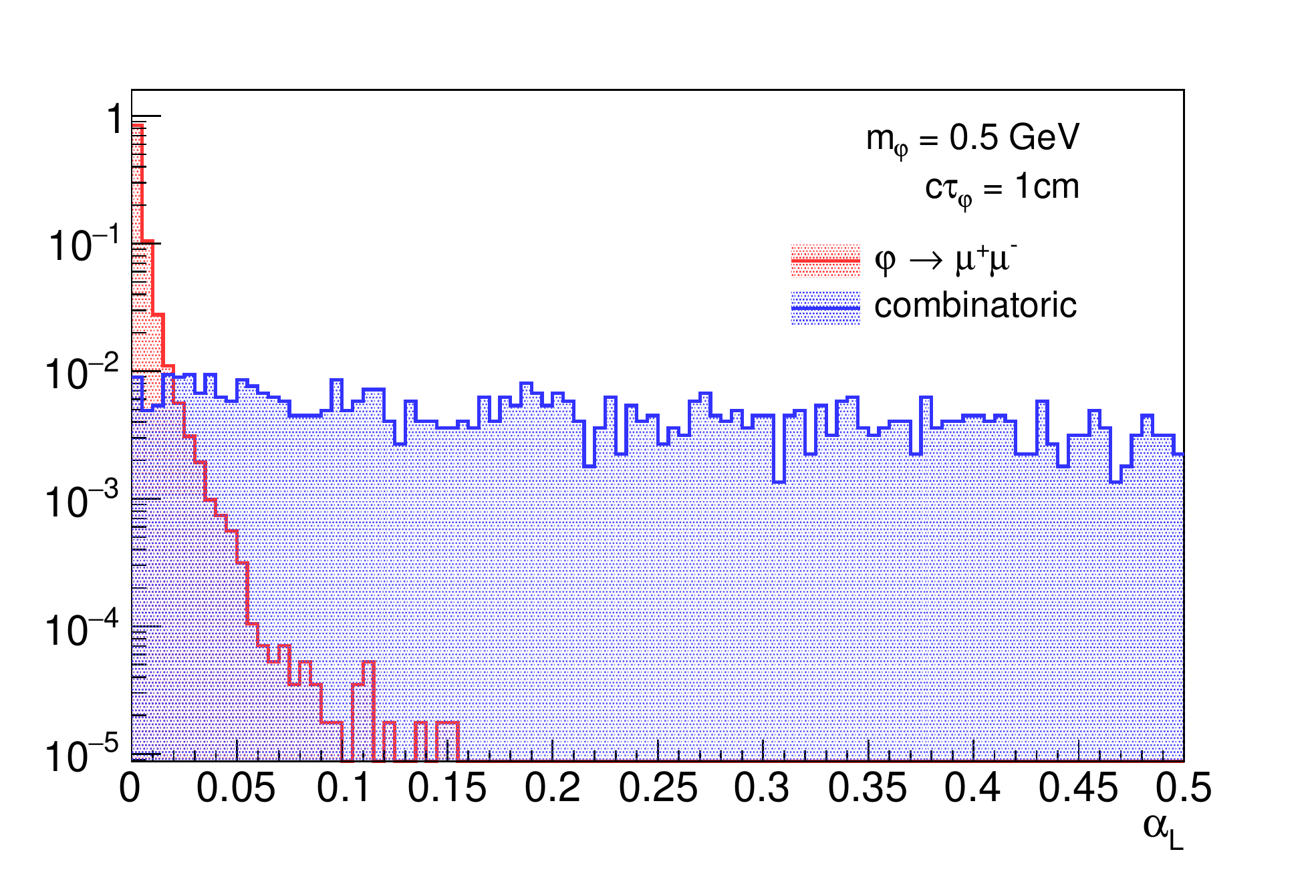}%
\caption{\label{fig:alpha_l} Angle between $\varphi$'s reconstructed momentum and the line connecting the vertex location with the origin, in $z$-$r$ plane,  for a signal with $m_\varphi=0.5$ GeV and $c\tau=1$ cm (red) and random crossings by fake tracks (blue).}
\end{figure}

Finally, with Fig.~\ref{fig:alpha_l} we also include the distribution of the impact angle in the longitudinal direction. If the primary vertex can be identified, cut on in this variable can further reduce the fake background to the extend that is negligible compared to $K_S$ decays.

\end{document}